\documentclass[letterpaper, 10 pt, conference]{ieeeconf}  

\IEEEoverridecommandlockouts                              
\usepackage[utf8]{inputenc}
\usepackage[T1]{fontenc}

\usepackage{amsmath,amssymb,amsfonts}
\usepackage{bm,bbm,siunitx,hyperref,graphicx}
\usepackage{float}	
\usepackage{xcolor}

\let\labelindent\relax

\usepackage{enumitem}
\usepackage{cite}
\usepackage{balance}

\title{\LARGE \bf
System Identification in Multi-Actuator Hard Disk Drives with Colored Noises using Observer/Kalman Filter Identification (OKID) Framework
}

\author{Nikhil Potu Surya Prakash$^{1}$ Zhi Chen$^{2}$ and Roberto Horowitz$^{3}$
\thanks{$^{1}$Nikhil Potu Surya Prakash is with Department of Mechanical Engineering,
        University of California, Berkeley, CA 94720, USA
        {\tt\small nikhilps@berkeley.edu}}%
\thanks{$^{2}$Zhi Chen is with Department of Mechanical Engineering,
        University of California, Berkeley, CA 94720, USA
        {\tt\small chenzhi@berkeley.edu}}%
\thanks{$^{3}$Roberto Horowitz is with Department of Mechanical Engineering,
        University of California, Berkeley, CA 94720, USA
        {\tt\small horowitz@berkeley.edu}}%
}

\begin{document}

\maketitle
\thispagestyle{empty}
\pagestyle{empty}

\begin{abstract}
Multi Actuator Technology in Hard Disk Drives (HDDs) equips drives with two dual stage actuators (DSA) each comprising of a voice coil motor (VCM) actuator and a piezoelectric micro actuator (MA) operating on the same pivot point. Each DSA is responsible for controlling half of the drive’s arms. As both the DSAs operate independently on the same pivot timber, the control forces and torques generated by one affect the operation of the other. The feedback controllers might not completely reject these transferred disturbances and a need to design feedforward controllers arises, which require a good model of the disturbance process. The usual system identification techniques produce a biased estimate because of the presence of the runout which is a colored noise. In this paper, we use the OKID framework to estimate this disturbance cross transfer function from the VCM control input of one DSA to the output of the other DSA from the collected time series data corrupted by colored noise.

\end{abstract}

\section{Introduction}
In 2017, Seagate unveiled its new multi actuator technology as a breakthrough that can double the data transfer performance of the future-generation hard drives for hyper-scale data centers. In a multi actuator drive, the read-write (R/W) heads are split into two sets, an upper and a lower half, which can double the data transfer rate by having the upper and lower platter sets work in parallel.

The multi actuator setup brought some new challenges to the HDD's controller design.
Since both the DSAs operate on the same pivot point, the forces and torques generated by
one DSA can affect the operation of the other DSA. The interaction of
the two DSAs can be categorized into three basic scenarios. In the first scenario, both
DSAs are in track following mode, and it is expected that the interaction between the
two actuators is negligible. In the second scenario, both the DSAs are in seek mode. In
this mode, the coupling vibrational interaction is usually negligible compared to the large
trajectories for both DSAs. In the third scenario, one DSA is in seek mode
and the other is in track following mode. Under this scenario, the seeking DSA
will impart disturbances in the form of vibration to the track following DSA, which
hamper the performance of the track following DSA drastically. The feedback controllers might not completely reject these transferred disturbances and a  need  to  design  feedforward  controllers  arises,  which  require a  good  model  of  the  disturbance  process. 

Modern HDDs use the inbuilt actuators to write the servo tracks instead of a servowriter or disk writer to reduce the cost of manufacturing. A servowriter uses big motors and robust equipment to write the servo tracks precisely which would be too bulky to be incorporated in a HDD. This amount of precision is not possible during the writing process by a HDD itself as the actuators are not as robust as a servowriter. The disturbances such as spindle vibrations, windage etc affect the writing process and hence the servo tracks are not exactly circles as desired. This offset is called the runout which can be modeled as white noise colored by some stable disturbance process. The presence of this runout corrupts the time series signals by a colored noise.   

System identification techniques like the series-parallel algorithm produce  a  biased  estimate even when the signal is corrupted by white noise. Parallel predictors though estimate an unbiased estimate, they suffer from stability issues and coming up with the necessary filters requires apriori knowledge of the system. In  this paper, we modify the Observer/Kalman filter Identification (OKID) technique to estimate this disturbance cross transfer function from the VCM control input of one DSA to  the  output  of  the  other  DSA  by using  the  collected  time  series data  corrupted  by  colored  noise.

OKID \cite{c1,c3,c5,c6,c7} is a method of identification of a linear dynamical
system along with the associated Kalman filter from input-output measurements corrupted by noise. OKID was originally developed at NASA as the OKID/ERA algorithm. Compared to other approaches, OKID is formulated via state observers providing an intuitive interpretation from a control theory perspective.
\section{Observer/Kalman filter Identification formulation}
In this section, we modify the OKID algorithm to account for colored noises. We formulate for the case when a colored noise is added to the output of the system as the runout in HDDs is colored, but the same technique can be used when the process noise is also colored. 
Consider the following dynamical system of interest
\begin{align}\label{actual system}\
\begin{split}
        x_a(k+1)=A_ax(k)+B_a u(k)+\bar{w}_p(k) \\
        y_a(k) = C_ax_a(k)+D_a u(k)+w_{cm}(k)
\end{split}
\end{align}
where $x_a(k)$ is the state of the system, $\bar{w}_p(k)$ is a white process noise, $u(k)$ is the known control input, $y_a(k)$ is the measured output, $w_{cm}$ is a colored noise (runout in the case of a HDD) and ($A_a,B_a,C_a,D_a$) represent the state, control, output and feedthrough matrices respectively of the actual system. The subscript 'a' is used for the actual system of interest.\\

The colored noise can be assumed to be the output of a stable dynamical system with a white noise as its input as follows
\begin{align}\label{coloring dynamics}
\begin{split}
        x_c(k+1)=A_cx(k)+w_{wm}(k) \\
    y_c(k) = C_cx_c(k)+D_c w_{wm}(k)
\end{split}
\end{align}
where $x_c(k)$ is the state of the coloring process, $w_{wm}(k)$ is a white noise, $y_c(k)$ is is the colored noise ($y_c(k) = w_{cm}(k)$) and ($A_c,B_c,C_c,D_c$) represent the state, control, output and feedthrough matrices respectively of the coloring process. The subscript 'c' is used for the coloring process.\\

Both the dynamical systems in \eqref{actual system} and \eqref{coloring dynamics} can be augmented to form
\begin{align}\label{eq:augmentedstatespace}
\begin{split}
       x(k+1)&=Ax(k)+Bu(k)+w_p(k) \\
         y(k) &= Cx(k)+Du(k)+w_m(k) 
\end{split}
\end{align}

where 
\begin{multline*}
x = 
    \begin{bmatrix}
        x_a \\
        x_c
    \end{bmatrix},
    A = 
        \begin{bmatrix}
        A_a & 0 \\
        0 & A_c
    \end{bmatrix},
    B = 
        \begin{bmatrix}
        B_a \\
        0
    \end{bmatrix},
    C = 
        \begin{bmatrix}
        C_a & C_c
    \end{bmatrix},\\
    D = D_a,
    w_p = 
        \begin{bmatrix}
        \bar{w}_p \\
        w_m
    \end{bmatrix} \; \& \;
    w_m = D_c w_{wm}
\end{multline*}

It can also be seen that the transfer function from the output to the input of the augmented system is the same as that of the actual system.
\begin{equation}
    \frac{Y_a(s)}{U(s)} = C_a(sI-A_a)^{-1}B_a+D_a
\end{equation}
\begin{equation}
\begin{split}
    \frac{Y(s)}{U(s)} &= C(sI-A)^{-1}B+D \\&= [C_a\;C_c](sI-\begin{bmatrix}
          A_a & 0\\
          0&A_c
    \end{bmatrix})^{-1} 
    \begin{bmatrix}
    B_a\\
    0
    \end{bmatrix}+D \\&= [C_a\;C_c]\begin{bmatrix}
          (sI_a-A_a)^{-1} & 0\\
          0&(sI_c-A_c)^{-1}
    \end{bmatrix}
    \begin{bmatrix}
    B_a\\
    0
    \end{bmatrix}+D\\
    &=C_a(sI_a-A_a)^{-1}B_a+D_a
\end{split}
\end{equation}
since $D=D_a$. Here $I,I_a$ and $I_c$ represent identity matrices with same sizes as $A, A_a$ and $A_c$ respectively.

Here neither the matrices, the noises nor their variances and covariances are assumed to be known. The only assumption made is that the noises in (\ref{eq:augmentedstatespace}) are zero mean gaussian.

Since the pair $(A_a,C_a)$ is detectable and $A_c$ is Schur, the pair $(A,C)$ will also be detectable if $C_c \neq 0$.Now, as $(A,C)$ is detectable, a steady state Kalman filter can be designed for the system if we exactly know the matrices $(A,B,C,D)$. Hence a Kalman filter gain $K$ exists such that $A-KC$ is Schur. With the filter gains defined (as unknowns), the observer dynamics can be expressed as
\begin{equation}\label{observer}
    \begin{split}
        \hat{x}(k+1) &= (A-KC)\hat{x}(k)+(B-KD)u(k)+Ky(k) \\
        \hat{y}(k) &= C\hat{x}(k)+Du(k)
    \end{split}
\end{equation}
Let $F = A-KC,\; H = B-KD$ and $G = K$\\
Defining $L = [H\;G]$ and $\nu_x(k) = [u^T(k)\;y^T(k)]^T$ for brevity, we get the observer in predictor form as
\begin{equation}\label{FL}
    \begin{split}
        \hat{x}(k+1) = F\hat{x}(k)+Lv_x(k) \\
        \hat{y}(k) = C\hat{x}(k)+Du(k)
    \end{split}
\end{equation}
Using \eqref{observer}, the observer state at $k^{th}$ time instance can be derived as
\begin{equation}\label{obsstateprg}
    \hat{x}(k) = F^p \hat{x}(k-p)+Tz(k)
\end{equation}

where 
\begin{align*}
    T &= 
    \begin{bmatrix}
    I & F & ... & F^{p-2} & F^{p-1}
    \end{bmatrix} L, \; \\
    z(k) &= 
    \begin{bmatrix}
    \nu^T_x(k-1) & \nu^T_x(k-2) &...& \nu^T_x(k-p)
    \end{bmatrix}
\end{align*}
The stability of the observer ensures that $F^p$ becomes negligible for sufficiently large values of  $p \; (p>>n)$ and hence the observer state in \eqref{obsstateprg} becomes
\begin{equation}\label{obsaprxstate}
    \hat{x}(k) \approx Tz(k)
\end{equation}
Substituting \eqref{obsaprxstate} in \eqref{FL}, we get the estimated output as
\begin{equation}\label{estmy}
    \hat{y}(k) = CTz(k)+Du(k)
\end{equation}
The estimated output in \eqref{estmy} is related to the measured output as
\begin{equation}
    y(k) = \Phi \nu(k)+\epsilon(k)
\end{equation}
where $\nu(k) = [u^T(k) z^T(k)]^T$, $\Phi = [D\;CL\;CFL\;...\;CF^{p-2}L\;CF^{p-1}L]$ and $\epsilon(k)$ is the error between the measured output and the estimated output.\\
The outputs at different time instances can be collected and stacked to obtain the following equation
\begin{equation}\label{Ystack}
    Y = \Phi V+E
\end{equation}
where
\begin{subequations}
    \begin{equation}
        Y = [y(p)\;y(p+1)\;...\;y(l-1)]
    \end{equation}
    \begin{equation}
        V = [\nu(p)\;\nu(p+1)\;...\;\nu(l-1)]
    \end{equation}
    \begin{equation}
        E = [\epsilon(p)\;\epsilon(p+1)\;...\;\epsilon(l-1)]
    \end{equation}
\end{subequations}
for $l$ measurements. The best estimate of $\Phi$ is obtained using least squares formulation as $\hat{\Phi} = YV^{\dagger}$ ($\dagger$ denotes the pseudo inverse). Various system matrices and the Kalman filter gain can be extracted from $\hat{\Phi}$ using the Eigensystem Realization Algorithm (ERA).

\section{Eigensystem Realization Algorithm (ERA)}
Eigensystem Realization Algorithm, first developed in \cite{c22}, is a system identification technique used most popularly for aerospace and civil structures from the input and output time domain data. Though ERA uses impulses as inputs to excite the system, it can be intertwined with OKID framework to estimate the system matrices using non-impulsive inputs. There are many variants to the ERA depending on the application and the type of data collected. In this section we will summarize one of the variants to extract estimates of the system matrices and the Kalman filter gain from $\hat{\Phi}$. 

It can be easily seen that $D$ is the first column in $\hat{\Phi}$. From the Markov parameters in $\hat{\Phi}$, the following Hankel matrices can be defined
\begin{equation}
\mathcal{H}(0) =
    \begin{bmatrix}
    CL & CFL & CF^2L & \dots\\
    CFL & CF^2L &CF^3L & \dots \\
    \vdots & \vdots &\ddots
    \end{bmatrix} = \mathcal{O}\mathcal{C}
\end{equation}

\begin{equation}
\mathcal{H}(1) = 
    \begin{bmatrix}
    CFL & CF^2L & CF^3L & \dots\\
    CF^2L & CF32L &CF^4L & \dots \\
    \vdots & \vdots &\ddots
    \end{bmatrix} = \mathcal{O}F\mathcal{C}
\end{equation}
where $\mathcal{O}$ is the observability matrix and $\mathcal{C}$ is the controllability matrix of the dynamical system in \eqref{FL} with $\nu_x$ as its input.\\
Using singular value decomposition, $\mathcal{H}(0)$ can be decomposed as
\begin{equation}
    \mathcal{H}(0) = U \Sigma V^T = 
    \begin{bmatrix}
          U_{+} & U_{-}
    \end{bmatrix}
    \begin{bmatrix}
          \Sigma_{+} & 0 \\
          0 & \Sigma_{-}
    \end{bmatrix}
    \begin{bmatrix}
          V^T_{+} \\ V^T_{-}
    \end{bmatrix}
\end{equation}
The subscript '$+$' denotes the singular values above a specified threshold and '$-$' for the singular values below the threshold. The desired degree of the estimated system can be decided from these dominant singular values.\\
The Observability and Controllability matrices can be split as follows (\cite{c23} presents other ways in which the matrices can be split).
\begin{equation}
\mathcal{O} = U_{+}\Sigma_{+}^{\frac{1}{2}} 
\end{equation}
\begin{equation}
\mathcal{C} = \Sigma_{+}^{\frac{1}{2}}V^T_{+} 
\end{equation}
Now using the Hankel matrix $\mathcal{H}(1)$, the matrix $F$ can be obtained as
\begin{align}
\mathcal{H}(1) &= \mathcal{O}F\mathcal{C} = U_{+}\Sigma_{+}^{\frac{1}{2}}F\Sigma_{+}^{\frac{1}{2}}V^T_{+} \nonumber \\ \implies F &=  \Sigma_{+}^{-\frac{1}{2}}U^T_{+}\mathcal{H}(1)V_{+}\Sigma_{+}^{-\frac{1}{2}}
\end{align}
Since $\mathcal{C}$ is the controllability matrix of \eqref{FL}, $L$ can be obtained from the first few columns of $\mathcal{C}$ and similarly $C$ can be obtained from the first few rows of $\mathcal{O}$ based on the dimensions. Further $H$ and $G$ can be obtained from $L$ as $L=[H\;G]$.
The Kalman filter gain $K=G$. Now, the state matrix $A$ and the input matrix $B$ can be estimated from $F$ and $H$ matrices respectively.\\
\begin{equation}
    F=A-KC \implies A=F+GC
\end{equation}
\begin{equation}
    H=B-KD \implies B=H+GD
\end{equation}

\section{Results}
The algorithm extended to include colored noises in OKID/ERA framework has been used to estimate the disturbance cross transfer function between the voice coil motor input of the track seeking actuator and the output of the track following actuator in a multi actuator hard disk drive. The simulation results will be presented in this section.

\begin{figure}[htpb]
    \centering
    \includegraphics[width=0.9\columnwidth]{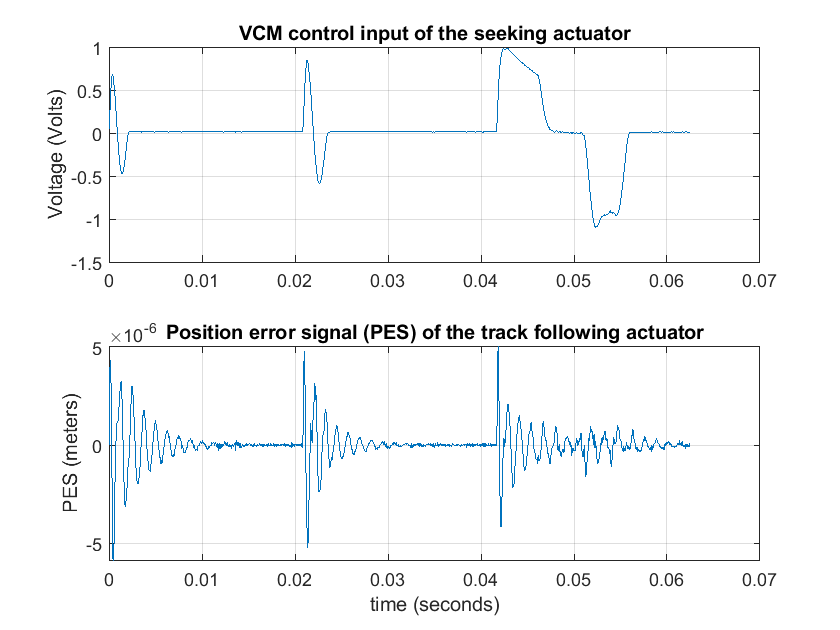}
    \caption{The input and output time series used for the simulations}
    \label{fig:InputandOutput}
\end{figure}

\begin{figure}[htpb]
    \centering
    \includegraphics[width=0.9\columnwidth]{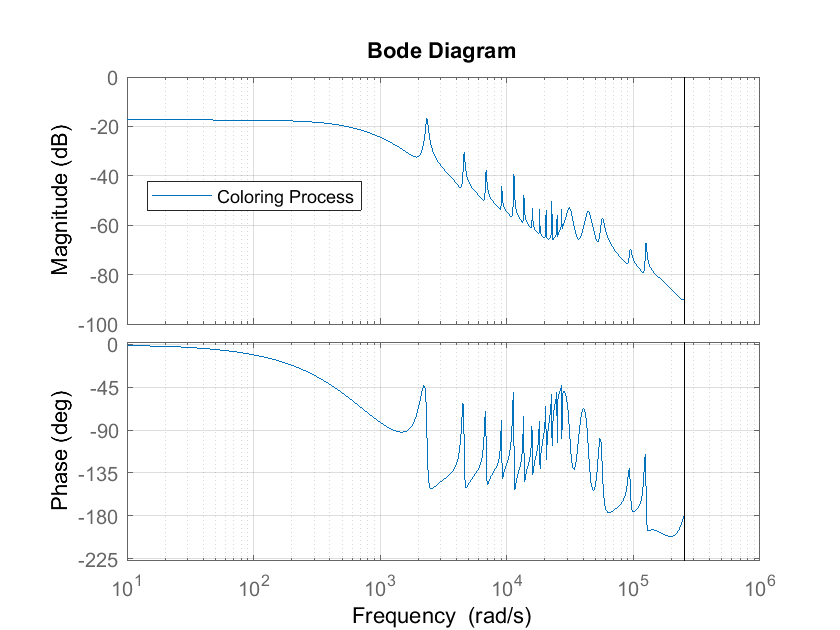}
    \caption{The frequency response plot of the coloring dynamics used for simulations}
    \label{fig:Runout}
\end{figure}

Fig.\ref{fig:InputandOutput} shows three seeking inputs one after the other in the top plot. This series forms the control input to the DSA in the track seeking mode. As both the DSAs in the multi actuator drive are mounted on the same pivot timber, the large movements of the track seeking actuator cause vibrations to transfer through the pivot timber and affect the track following DSA. For this estimation, the track following DSA is turned off so that its free response can be used to estimate the cross transfer function. The position error signal (PES) is collected during the excitation and used for the estimation.The sampling frequency in this case was 38520 Hz. The bottom plot of fig. \ref{fig:InputandOutput} shows the PES as a function of time.  Fig.\ref{fig:Runout} shows the frequency response plot of the coloring process.

\begin{figure}[htpb]
    \centering
    \includegraphics[width=0.9\columnwidth]{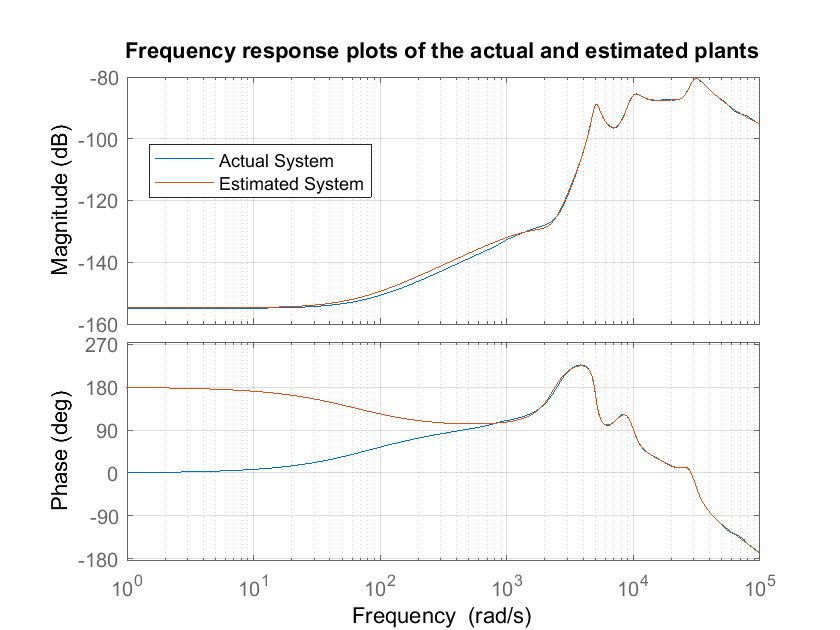}
    \caption{Frequency responses of the actual plant and the estimated plant}
    \label{fig:CTFbode}
\end{figure}

 A total of 1600 measurements were used and $p$ was chosen to be 800. Ten most dominant singular values of the Hankel matrix were used for the estimation which gave a tenth order plant estimate. Fig.\ref{fig:CTFbode} shows frequency response plots of the actual and estimated plots. The PES collected had more high frequency components and hence a more accurate fit in the high frequency region was obtained.

\section{Conclusions}\label{sec:Conc}
In this paper the OKID/ERA algorithm was extended to include colored noises. The algorithm was used to estimate the disturbance cross transfer function between the voice coil motor input of the track seeking actuator and the output of the track following actuator in a multi actuator hard disk drive.
\section*{ACKNOWLEDGMENT}
We would like to acknowledge the feedback and financial support from ASRC hosted by the International Disk Drive Equipment and Materials Association (IDEMA).

\end{document}